\documentclass[twocolumn,preprintnumbers,amsmath,superscriptaddress,amssymb,aps]{revtex4-1}

\usepackage{graphicx}
\usepackage{dcolumn}
\usepackage{bm}
\usepackage{amsmath}
\usepackage{amssymb}
\usepackage{graphicx}
\usepackage{indentfirst}
\usepackage{booktabs}
\usepackage{multirow}
\usepackage{colortbl}

\selectfont

\begin{document}

\title{Strong bulk photovoltaic effect and second-harmonic generation in two-dimensional selenium and tellurium}

\author{Meijuan Cheng}
\address{Department of Physics, Collaborative Innovation Center for Optoelectronic Semiconductors
and Efficient Devices, Key Laboratory of Low Dimensional Condensed Matter Physics
(Department of Education of Fujian Province), Jiujiang Research Institute, Xiamen University, Xiamen 361005, China}
\author{Zi-Zhong Zhu}
\email{zzhu@xmu.edu.cn}
\address{Department of Physics, Collaborative Innovation Center for Optoelectronic Semiconductors
and Efficient Devices, Key Laboratory of Low Dimensional Condensed Matter Physics
(Department of Education of Fujian Province), Jiujiang Research Institute, Xiamen University, Xiamen 361005, China}
\address{Fujian Provincial Key Laboratory of Theoretical and Computational Chemistry, Xiamen 361005, China}
\author{Guang-Yu Guo}
\email{gyguo@phys.ntu.edu.tw}
\address{Department of Physics and Center for Theoretical Physics, National Taiwan University, Taipei 10617, Taiwan}
\address{Physics Division, National Center for Theoretical Sciences, Taipei 10617, Taiwan}


\date{\today}

\begin{abstract}
Few-layer selenium and tellurium films have been recently prepared, and 
they provide a new platform to explore novel properties of two-dimensional (2D) elemental materials. 
In this work, we have performed a systematic first-principles study 
on the electronic, linear and nonlinear optical (NLO) properties of atomically thin selenium and tellurium films within
the density-functional theory with the generalized gradient approximation plus scissors correction using
the band gaps from the relativistic hybrid Heyd-Scuseria-Erzerhof functional calculations. 
The underlying atomic structures of these materials are determined theoretically 
using the SCAN exchange-correlation functional.
Interestingly, we find that few-layer
Se and Te possess large second-harmonic generation (SHG), linear electro-optic (LEO) effect and bulk photovoltaic effect. 
In particular, trilayer (TL) Te possesses large SHG coefficient, being more than 65 times larger than that of GaN, a widely used NLO material.
Bilayer (BL) Te has huge static SHG coefficient $\chi^{(2)}_{xyy}(0)$, being more than 100 times larger than that of GaN. 
Furthermore, monolayer (ML) Se possesses large SHG coefficient with $\chi^{(2)}_{xyy}$ 
being six times larger than that of GaN. Both ML Se and BL Te possess large linear
electro-optic coefficients $r_{xyy}(0)$ and $r_{yzx}(0)$, which is about 6 times and 5 times larger than that of bulk GaN polytypes, respectively. 
Moreover, we predict that TL Te exhibits strong bulk photovoltaic effect (BPVE) with shift current conductivity 
of $\sim$440 $\mu$A/V$^2$, being greater than that of GeS, a polar system with the largest BPVE found so far.
Although the shift current conductivities of bulk and 2D Se are comparable,
the shift current conductivities of TL Te are five times larger than that of bulk Te. 
Finally, an analysis of the calculated electronic band structures indicates that the strong NLO responses 
of 2D Se and Te materials are primarily derived from their low-dimensional structures with high anisotropy, 
directional covalent bonding, lone-pair electrons and relatively small band gaps. These findings provide
a practical strategy to search for excellent NLO and BPVE materials.
\end{abstract}

\maketitle


\section{INTRODUCTION}
Noncentrosymmetric materials under intense optical fields may generate large even-order nonlinear optical (NLO) responses 
such as second-order NLO susceptibility ($\chi^{(2)}$)~\cite{shen1984,Boyd2003}.
Large second-order NLO susceptibility is of great importance for many technological 
applications such as electro-optical switches, light signal modulators and frequency conversion.
As one of the best-known second-order NLO optical responses \cite{Boyd2003}, the second-harmonic generation (SHG) 
has been widely used as surface probes and frequency doublers \cite{shen1984}. 
Since the 1960s, the SHG has been investigated extensively in
bulk semiconductors \cite{Boyd2003,Chang1965,zhong1993,Hughes1996,Gavrilenko2000,Cai2009,Cheng2019} and more recently also in
one-dimensional (see, e.g., Refs. \cite{guo2004,guo2005} and references therein) and two-dimensional (see, e.g., 
Refs. \cite{Gruning2014,Trolle2014,wang2015,Attaccalite2015,hu2017,wang2017,panday2018,Attaccalite2019} and references therein) materials. 
Linear electro-optic (LEO) effect, another second-order NLO response of a noncentrosymmetric material,
refers to the linear refractive index change ($\Delta n$) with the applied electric field strength ($E$), $\Delta n = n^3rE/2$, where $n$ is
the refraction index and $r$ is the LEO coefficient \cite{Boyd2003}.
The LEO effect thus allows one to use an electrical signal to control the amplitude, phase or direction of a light beam in the NLO material,
and leads to a widely used means for high-speed optical modulation and sensing devices (see, e.g., Ref. \cite{Wu1996} and references therein).

In recent years we have seen a surge of interest in bulk photovoltaic effect (BPVE), another nonlinear optical response, referring to the generation
of photovoltage or photocurrent in noncentrosymmetric materials. Earlier studies in 1960s demonstrated intrinsic photocurrents in ferroelectric oxide
BaTiO$_3$ \cite{Chynoweth1956}. Afterwards, the BPVE was reported in LiNbO$_3$ \cite{Glass1974}. 
Recently, Young {\it et al.} demonstrated theoretically that shift photocurrent dominates the BPVE in BaTiO$_3$.~\cite{Young2012} 
Then, Bhatnagar {\it et al.} experimentally
discovered large open-circuit photovoltages and thus proved the BPVE in multiferroic BiFeO$_3$ (BFO) thin films.~\cite{Bhatnagar2013} 
After this, Brehm {\it et al.} calculated the shift current conductivity for polar compounds LiAsS$_2$, LiAsSe$_2$ and NaAsSe$_2$, 
and found that they exhibit shift current which is nearly 20 times larger than that of BiFeO$_3$ \cite{Brehm2014}.
Recently, Rangel {\it et al.} theoretically investigated the BPVE in single-layer monochalcogenides \cite{Rangel2017}. 
More recently, Gong {\it et al.} found that ferroelectric semiconductor GeTe possesses large shift current
response due to narrow band gap and high covalency \cite{Gong2018}. 

Selenium and tellurium bulks \cite{agapito2013,peng2014,Hirayama2015,csahin2018,tsirkin2018} and
chains \cite{Churchill2017,Li2005,Olechna1965,Springborg1988,Ghosh2007,Kahaly2008,Tuttle2017,Andharia2018,Pan2018} 
have been investigated extensively due to their unique properties. Nevertheless, their two-dimensional (2D) counterparts
have not been fabricated and investigated almost untill the present day. In 2017, Qin {\it et al.} first achieved 
the controlled growth of a large-size 2D selenium nanosheet using a physical
vapor deposition method \cite{Qin2017}. Zhu {\it et al.} revealed that the metalloid element Te exists three phases of 2D monolayer,
1T-MoS$_2$-like ($\alpha$-Te) structure, tetragonal ($\beta$-Te) and 2H-MoS$_2$-like ($\gamma$-Te) structures, named tellurene \cite{Zhu2017}.
In particular, $\beta$-Te has been synthesized on highly oriented pyrolytic graphite (HOPG) substrates by using molecular
beam epitaxy \cite{Zhu2017}. Since then, extensive investigations on 2D Se and Te
materials have been conducted both theoretically \cite{Xian2017, Xue2017, Sharma2018, Debela2018, Zhang2018, Chen2019, Liu2019} and
experimentally \cite{Chen2017, Wang2018, Peng2018, Apte2018, Kang2019, Deckoff-Jones2019}.
Notably, Huang {\it et al.} recently grew monolayer (ML) and few-layer Te films on a graphene/6H-SiC(0001) substrate by molecular beam epitaxy.
The Te films  consist of the parallel helical chains exposing the Te bulk $b-c$ (or $a-c$) facet \cite{Huang2017}. 
Remarkably, these ML and few-layer films are noncentrosymmetric and thus are potentially excellent 2D second-order NLO materials.
Soon afterwards, several theoretical groups investigated the structural, electronic, optical properties \cite{qiao2018}, ferroelectricity and
spin-textures \cite{Wangy2018}, as well as phase transitions \cite{Wangc2018} in few-layer tellurium films.
Strong optical absorption, high carrier mobility,  outstanding environmental stability, intrinsic anisotropy, 
layer-dependent and almost direct band gap of these films have been observed, 
indicating their promising applications in electronics, optoelectronics, vapor sensors, spintronics, biomedicine.

Stimulated by these experimental breakthroughs, here we have carried out a systematic first-principles investigation on the SHG
and  BPVE in these 2D selenium and tellurium based on the density functional theory calculations.
This paper is organized as follows.  In Sec. \uppercase\expandafter{\romannumeral2}, we present the theoretical approach and computational details.
In Sec. \uppercase\expandafter{\romannumeral3}, we report the electronic band structures of 2D $\alpha$-phase Se and Te. 
In Sec. \uppercase\expandafter{\romannumeral4}, we present the real and imaginary parts of the optical dielectric function. 
In Sec. \uppercase\expandafter{\romannumeral5}, the calculated LEO and SHG coefficients over the entire optical
frequency range are presented. Furthermore, the second-order optical susceptibility is analyzed in terms of one- and two-photon resonances.
In Sec. \uppercase\expandafter{\romannumeral6}, we present the calculated shift current conductivities. 
Comparison of the obtained results of 2D Se and Te with other known NLO materials suggests that they  
have promising applications in NLO devices such
as novel solar-cell, photodetector, SHG, electro-optical modulator and electric optical switches. 
In Sec. \uppercase\expandafter{\romannumeral7}, we discuss the possible origins of the strong NLO responses of 2D Se and Te
in terms of the calculated deformation charge density distribution. Finally, a summary is
given in Sec. \uppercase\expandafter{\romannumeral8}.

\section{STRUCTURE AND COMPUTATIONAL METHOD}

The $\alpha$-phase crystalline structure of bulk selenium and tellurium is schematically shown in Figs. 1(a), 1(b) and 1(c). 
It consists of parallel-aligned helical chains arranged in a hexagonal array \cite{Teuchert1975,Keller1977}. 
Every atom is covalently bonded to two neighboring
atoms along each chain, and interacts weakly with atoms on the adjacent chains. 
Figure 1(a) shows that bulk selenium and tellurium 
may be viewed as a layered structure. Different layers are stacked in the direction perpendicular to $a-c$ (or $b-c$) planes 
and each layer consists of one-dimensional array of single helical chains stretched to infinity along the $c$-axis. 
An $n$ layer 2D structure is made of $n$ helical chain layers stacked in the bulk mode [see Figs. 1(d), 1(e) and 1(f)].  
Interestingly, the space group of an odd-number layer structure is $P2$ ($C_2^1$), 
while that of an even-number layer structure is $P2_1$ ($C_2^2$) (see Table II). 
Note that $\alpha$-phase bulk and few-layer structures all possess broken inversion symmetry, 
the necessary condition for nonzero second-order NLO responses. However, the $\alpha$-phase ML tellurium
is unstable against the $\beta$-phase, a structure with centrosymmetry [see Figs. 1(g), 1(h) and 1(i)]. 
Nonetheless, for all other atomically thin Te films, $\alpha$-phase is the most stable structure.
Also, for all the atomically thin Se films, $\alpha$-phase is the most stable structure.
Therefore, we consider only bilayer (BL) and trilayer (TL) tellurium as well as ML, BL and TL selenium.  
In the present calculations, 2D selenium and tellurium structures are modeled 
by using the slab-superlattice approach. The separations
between the 2D structures adopted are about 20 \AA, to ensure negligible interaction between the periodic images.

\begin{figure*}[htb]
\begin{center}
\includegraphics[width=14cm]{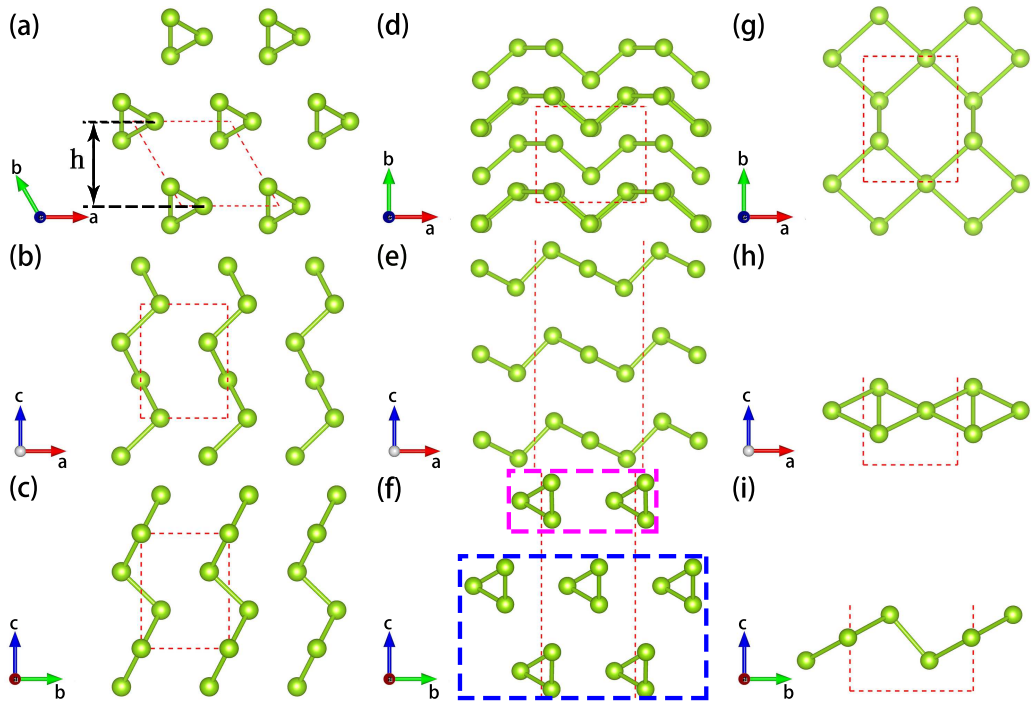}
\end{center}
\caption{(a) Top and (b,c) side views of the $\alpha$-phase crystalline structure of bulk selenium and tellurium.
(d) Top and  (e, f) side views of the $\alpha$-phase TL structure.
In (a), symbol $h$ denotes the effective ML thickness.
In (f), the blue dashed rectangle indicates the $\alpha$-phase BL structure.
The magenta dashed rectangle indicates the $\alpha$-phase ML structure.
(g) Top and (h, i) side views of the $\beta$-phase ML Te. The red dashed lines indicate the unit cell of each structure.}
\end{figure*}

\begin{table}[htbp]
\begin{center}
\caption{Theoretical lattice constants (in the unit of \AA) of bulk selenium and tellurium
calculated with the SCAN, optB88-vdW, DFT-D2 and PBE functionals,
compared with the experimental lattice constants (exp).
}
\begin{ruledtabular}
\begin{tabular}{ccccccc}
     &     &SCAN & optB88-vdW & PBE-D2 & PBE & exp    \\
\hline
  Se & $a$ & 4.4336 & 4.2348 & 4.2517 & 4.4984 & 4.3662\footnotemark[1]  \\
     & $c$ & 4.9984 & 5.1319 & 5.1065 & 5.0526 & 4.9536\footnotemark[1]  \\
  Te & $a$ & 4.4578 & 4.4816 & 4.3357 & 4.5085 & 4.4511\footnotemark[2]  \\
     & $c$ & 5.9291 & 5.9951 & 6.0300 & 5.9593 & 5.9262\footnotemark[2]  \\
\end{tabular}
\end{ruledtabular}
\footnotemark[1]{Experimental value from reference~\onlinecite{Teuchert1975}.}

\footnotemark[2]{Experimental value from reference~\onlinecite{Keller1977}.}
\end{center}
\end{table}

\begin{table*}[htbp]
\begin{center}
\caption{Space group, calculated lattice constants ($a$, $b$ and $c$), effective thickness ($h$),  
layer-averaged interchain ($R$) and intrachain ($r$) distances as well as their ratio ($R/r$)
of bulk and 2D selenium and tellurium obtained using the SCAN functional.
}
\begin{ruledtabular}
\begin{tabular}{c c c c c c c c}
               &  bulk Se  & ML Se     & BL Se     & TL Se         &  bulk Te   & BL Te     & TL Te     \\  \hline
  space group  &  $P3_1$21 & $P$2      & $P2_1$    & $P$2          & $P3_1$21   & $P2_1$    & $P$2      \\
   $a$ (\AA)   & 4.434     & 4.918     & 4.955     & 4.968         & 4.458      & 5.791     & 5.839     \\
   $b$ (\AA)   & 4.434     & 4.198     & 4.287     & 4.336         & 4.458      & 4.311     & 4.367     \\
   $c$ (\AA)   & 4.998     &           &           &               & 5.929      &           &           \\
   $h$ (\AA)   &           & 3.781     & 7.562     & 11.344        &            & 7.71      & 11.564    \\
   $R$ (\AA)   & 3.50      & 3.34      & 3.55      & 3.54          & 3.47       & 3.51      & 3.48     \\
   $r$ (\AA)   & 2.38      & 2.36      & 2.37      & 2.37          & 3.17       & 2.85      & 2.84     \\
   $R/r$       & 1.47      & 1.42      & 1.50      & 1.49          & 1.09       & 1.23      & 1.23     
\end{tabular}
\end{ruledtabular}

\end{center}
\end{table*}

Our density functional theory calculations are performed 
by using the highly accurate projector augmented wave (PAW) method \cite{blochl1994}, as implemented in
the VASP package \cite{kresse1996ab,kresse1996}. 
A large plane-wave cutoff energy of 450 eV is used throughout. 
The PAW potentials for Se $4s^2 4p^4$ and for Te $5s^2 5p^4$ are adopted to describe the electron-ion interaction.
The theoretical atomic positions and lattice constants are obtained when the forces acting on all the atoms are less
than 0.001 eV/\AA$ $ and the stresses are less 1.0 kBar. The total energy convergence criterion for the self-consistent
electronic structure calculations is $10^{-6}$ eV. The accurate tetrahedron method \cite{blochl1994tetrahedron} 
is used for the Brillouin zone integration. The self-consistent charge density calculations are performed 
with $k$-point meshes of 18$\times$20$\times$1 for 2D Se and 15$\times$20$\times$1 for multilayer Te. 

To obtain accurate structure parameters, we first perform structural optimizations for bulk selenium and tellurium
with the conjugate gradient technique using the SCAN \cite{Sun2015,Peng2016}, optB88-vdW \cite{Thonhauser2007},
Perdew-Burke-Ernzerhof (PBE) \cite{perdew1996}, DFT-D2 \cite{Grimme2006} exchange-correlation functionals. 
The calculated lattice constants are listed in Table I. The lattice constants calculated using the SCAN functional 
are in the best agreement with experiments \cite{Teuchert1975,Keller1977} (see Table I). 
Therefore, in all the geometric optimization calculations for the 2D structures, we adopt the SCAN functional.

Using the structural parameters for the 2D structures determined by using the SCAN functional, 
we first perform the self-consistent band structure calculations within the generalized gradient approximation
(GGA) of the PBE functional~\cite{perdew1996}. We then calculate the optical dielectric function and NLO responses of the 2D structures
from the calculated band structures within the linear response formalism with the independent-particle approximation. 
In particular, the imaginary part of the dielectric function $\varepsilon(\omega)$ due to direct
interband transition in the atomic units is given by (see, e.g., Refs. \cite{guo2004,Guo2005ab}),
\begin{equation}
\varepsilon_{a}'' (\omega) = \frac{4\pi^2}{\Omega\omega^2}
\sum_{i\in VB,j\in CB}\sum_{{\bf k}}w_{{\bf k}}|p_{ij}^{a}|^{2}
\delta(\epsilon_{{\bf k}j}-\epsilon_{{\bf k}i}-\omega),
\end{equation}
where $\Omega$  is the unit-cell volume, $w_{{\bf k}}$ is the weight for the ${\bf k}$ point 
and $\omega$  is  the photon energy. VB and CB stand for the valence and conduction bands, respectively.
The dipole transition matrix elements $p_{ij}^{a} = \langle\textbf{k}\emph{j}|\hat{p}_{a}|\textbf{k}i\rangle$, 
where $\hat{p}_{a}$ represents Cartesian component $a$
of the dipole operator, are obtained from the band structures within the PAW formalism \cite{Adolph2001}.
Here $|${\bf k}$n\rangle$ is the $n$th Bloch state function with momentum {\bf k}.
The real part of the dielectric function is then obtained from the calculated $\varepsilon''(\omega)$
by the Kramer-Kronig transformation \cite{guo2004,Guo2005ab},
\begin{equation}
\varepsilon'(\omega) = 1+\frac{2}{\pi}{\bf P} \int _{0}^{\infty }
d\omega ^{'}\frac{\omega '\varepsilon'' (\omega ')}{\omega^{'2}-\omega ^{2}},
\end{equation}
where ${\bf P}$ denotes the principal value of the integral.

The imaginary part of the second-order NLO susceptibility due to direct interband transitions is given by \cite{guo2004,guo2005}
\begin{equation}
\chi''^{(2)}_{abc}(-2\omega,\omega,\omega) = \chi''^{(2)}_{abc,VE}(-2\omega,\omega,\omega)+\chi''^{(2)}_{abc,VH}(-2\omega,\omega,\omega),
\end{equation}
where the contribution due to the so-called virtual-electron (VE) process is \cite{guo2004,guo2005}
\begin{equation}
\begin{split}
\chi''^{(2)}_{abc,VE} = -\frac{\pi}{2\Omega}\sum_{i\in VB}\sum_{j,l\in CB}\sum_{\bf k}w_{\bf k}
\{\frac{Im[p_{jl}^{a}\langle p_{li}^{b} p_{ij}^{c}\rangle]}{\epsilon_{li}^{3}(\epsilon_{li}+\epsilon_{ji})}\delta(\epsilon_{li}-\omega) \\
-\frac{Im[p_{ij}^{a}\langle p_{jl}^{b} p_{li}^{c}\rangle]}{\epsilon_{li}^{3}(2\epsilon_{li}-\epsilon_{ji})}\delta(\epsilon_{li}-\omega)
+\frac{16Im[p_{ij}^{a}\langle p_{jl}^{b} p_{li}^{c}\rangle]}{\epsilon_{ji}^{3}(2\epsilon_{li}^{3}-\epsilon_{ji}^{3})}\delta(\epsilon_{ji}-2\omega)\},
\end{split}
\end{equation}
and that due to the virtual-hole (VH) process is \cite{guo2004,guo2005}
\begin{equation}
\begin{split}
\chi''^{(2)}_{abc,VH} = \frac{\pi}{2\Omega}\sum_{i,l\in VB}\sum_{j\in CB}\sum_{\bf k}w_{\bf k}
\{\frac{Im[p_{li}^{a}\langle p_{ij}^{b} p_{jl}^{c}\rangle]}{\epsilon_{jl}^{3}(\epsilon_{jl}+\epsilon_{ji})}\delta(\epsilon_{jl}-\omega) \\
-\frac{Im[p_{ij}^{a}\langle p_{jl}^{b} p_{li}^{c}\rangle]}{\epsilon_{jl}^{3}(2\epsilon_{jl}-\epsilon_{ji})}\delta(\epsilon_{jl}-\omega)
+\frac{16Im[p_{ij}^{a}\langle p_{jl}^{b} p_{li}^{c}\rangle]}{\epsilon_{ji}^{3}(2\epsilon_{jl}-\epsilon_{ji})}\delta(\epsilon_{ji}-2\omega)\}.
\end{split}
\end{equation}
Here $\epsilon_{ji}$ = $\epsilon_{{\bf k}j}$-$\epsilon_{{\bf k}i}$ and $\langle p_{jl}^{b}p_{li}^{c}\rangle = \frac{1}{2}(p_{jl}^{b}p_{li}^{c}+p_{li}^{b}p_{jl}^{c})$.
The real part of the second-order NLO susceptibility is obtained from the calculated $\chi''^{(2)}_{abc}$
by the Kramer-Kronig transformation \cite{guo2004,guo2005}
\begin{equation}
\chi'^{(2)}(-2\omega,\omega,\omega) = \frac{2}{\pi}{\bf P} \int_0^{\infty}d\omega'
\frac{\omega'\chi''^{(2)}(2\omega',\omega',\omega')}{\omega'^2-\omega^2}.
\end{equation}
The LEO coefficient $r_{abc}(\omega)$ is related to the second-order NLO
susceptibility $\chi_{abc}^{(2)}(-\omega,\omega,0)$ \cite{Hughes1996,Prussel2018}.
In the zero frequency limit~\cite{Hughes1996,Prussel2018},
\begin{equation}
r_{abc}(0)=-\frac{2}{\varepsilon_a(0)\varepsilon_b(0)}\lim_{\omega\rightarrow 0}\chi_{abc}^{(2)}(-2\omega,\omega,\omega).
\end{equation}
In the very low frequency region, i.e. the photon energy $\hbar\omega$ well below the band gap,
$\chi_{abc}^{(2)}(-2\omega,\omega,\omega)$ and $n(\omega)$ are nearly constant. In this case, the LEO coefficient
$r_{abc}(\omega)\approx r_{abc}(0)$ \cite{wang2015,guo2004}.

The DC shift current along the $a$-axis is given by 
\begin{equation}
J_{a}=\sum_{bc}\sigma_{abc}(0;\omega,-\omega)E_b(\omega)E_c(-\omega),
\end{equation}
where $\sigma_{abc}(0;\omega,-\omega)$ is the third-rank conductivity tensor.~\cite{Sipe2000}
Within the length gauge formalism, $\sigma_{abc}(0;\omega,-\omega)$ can be written in terms of
the interband position matrix element $r_{ij}^a$ and its momentum derivative $r_{ij;b}^a$ 
(see Eq. (57) in Ref. ~\cite{Sipe2000}). By replacing $r_{ij}^a$ with $p_{ij}^a/i\epsilon_{ij}$,
one would obtain 
\begin{equation}
\begin{split}
\sigma_{abc}(0;\omega,-\omega)=\frac{{\pi}}{\Omega}\sum_{i \in VB}\sum_{j\in CB}\frac{1}{\epsilon_{ij}^2}\sum_{\bf k} w_{\bf k} \\
\sum_{l \ne i,j}Im\{\frac{p_{il}^a \langle p_{ji}^{b} p_{lj}^{c}\rangle}{2\epsilon_{il}} +
\frac{p_{lj}^a \langle p_{ji}^{b} p_{il}^{c}\rangle}{2\epsilon_{jl}} \} \delta(\epsilon_{ji}-\omega).
\end{split}
\end{equation}

In the present calculations, the function $\delta$ in Eqs. (1), (4), (5) and (9) is approximated by a Gaussian function with $\Gamma=0.2$ eV.
The ${\bf k}$ point weight $w_{{\bf k}}$ is set to $(1/N_k)$ where $N_k$ is the total number of the sampled $k$-points.
To obtain accurate optical properties, we perform calculations for selenium and tellurium with several different $k$-point meshes
until the calculated optical properties converge to a few percent. Therefore, adequately dense $k$-point meshes of 52$\times$60$\times$1
and 45$\times$60$\times$1 are adopted for selenium and tellurium, respectively.
Furthermore, 30 energy bands per atom are included in the present optical calculations in order to ensure that $\varepsilon'$
and $\chi'^{(2)}$ obtained by the Kramer-Kronig transformation are reliable.
In the slab-supercell approach, the unit cell volume $\Omega$ in Eqs. (1), (4), (5) and (9) 
is not well-defined for 2D Se and Te. 
In the literature, two ways have been adopted to resolve this 
problem.~\cite{Gruning2014,Trolle2014,wang2015,Attaccalite2015,hu2017,wang2017,panday2018,Attaccalite2019,Rangel2017}
One way is to present the 2D response functions in terms of per unit width simply by multiplying the
calculated quantities with the supercell lattice constant along the perpendicular direction~\cite{Trolle2014,wang2017}. 
Physically, this means the optical responses coming from a single atomically thin film~\cite{Trolle2014,wang2017}.
The other way is to use the effective unit cell 
volume~\cite{Gruning2014,wang2015,Attaccalite2015,hu2017,panday2018,Attaccalite2019,Rangel2017}, which is given by the area of
the in-plane unit cell times the effective thickness $h$ (see Table II).
This quasi bulk approach would allow one to compare the 
calculated quantities with that of bulk materials, and thus is adopted in this paper.
Nevertheless, one should keep in mind that to evaluate the total responses from a single film, 
one should multiply the quantities by the effective thickness $h$ rather than light penetration depth.

\section{Electronic band structure}

\begin{figure*}[htb]
\begin{center}
\includegraphics[width=14.0cm]{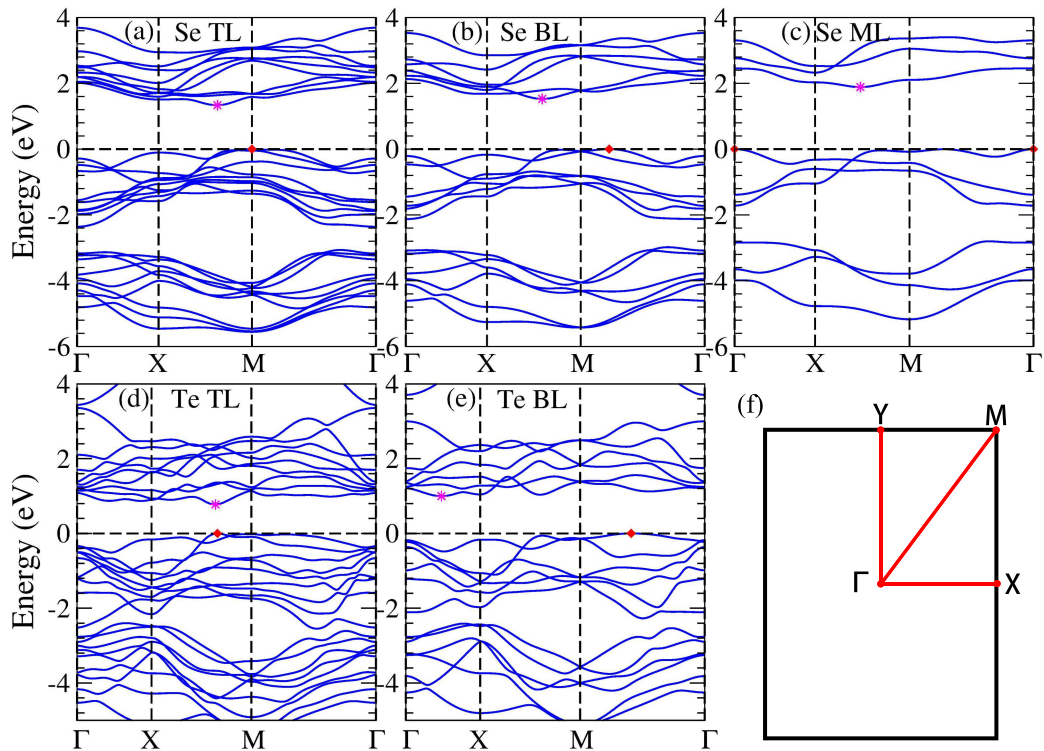}
\end{center}
\caption{Band structures of 2D (a, b, c) selenium as well as (d, e) tellurium. The valence band maximum (VBM) is marked as the red diamond,
while the conduction band minimum is indicated by the magenta star. (f) 2D Brillouin zone. The VBM is at 0 eV.}
\end{figure*}

\begin{table}[htbp]
\begin{center}
\caption{
Calculated ($E_{g}^{GGA}$, $E_{g}^{HSE-SOC}$) and experimental ($E_{g}^{Exp}$) band gaps
as well as scissors operator ($\Delta E_g = E_{g}^{HSE-SOC} - E_{g}^{GGA}$) for bulk and 2D selenium and tellurium.
}
\begin{ruledtabular}
\begin{tabular}{ccccccc}
    & &$E_{g}^{GGA}$ (eV)&$E_{g}^{HSE-SOC}$ (eV) &$E_{g}^{Exp}$ (eV)& $\Delta E_g$ (eV)   \\
\hline
  Se& ML  & 1.857 &  2.600      &                          &  0.756    \\
    & BL  & 1.503 &  2.270      &                          &  0.744     \\
    & TL  & 1.305 &  2.078      &                          &  0.743      \\
    &bulk & 1.002\footnotemark[1] &  1.735\footnotemark[1]  &  2.0\footnotemark[2] &    \\
  Te& BL  & 0.864 &1.128 (1.17\footnotemark[3]) &  0.85\footnotemark[4]     &  0.289    \\
    & TL  & 0.654 &1.072 (0.95\footnotemark[3]) &  0.74\footnotemark[4]    &  0.299    \\
    &bulk & 0.113\footnotemark[1] &0.322\footnotemark[1]  &  0.323\footnotemark[5]    &           \\
\end{tabular}
\end{ruledtabular}
\footnotemark[1]{Theoretical value from reference~\onlinecite{Cheng2019}.}

\footnotemark[2]{Experimental value from reference~\onlinecite{Tutihasi1967}.}

\footnotemark[3]{Theoretical value from reference~\onlinecite{qiao2018}.}

\footnotemark[4]{Experimental value from reference~\onlinecite{Huang2017}.}

\footnotemark[2]{Experimental value from reference~\onlinecite{Anzin1977}.}
\end{center}
\end{table}

To understand the electronic and optical properties of 2D Se and Te materials, we plot their electronic band structures in Fig. 2. 
Clearly, Fig. 2 shows that all the 2D Se and BL Te are indirect band gap semiconductors, 
while TL Te has a nearly direct band gap. For BL tellurium,
the valence band maximum (VBM) is close to the M point along the M-$\Gamma$ direction, while the conduction band minimum (CBM) is located somewhere
close to the $\Gamma$ along the $\Gamma$-X symmetry line. If the thickness of ultrathin Te film increases to that of TL, both VBM and
CBM are located at a general $k$-point close to the M along the X-M direction, i.e., TL Te is a nearly direct band gap
semiconductor and thus may have promising applications in semiconductor devices. 
In contrast, a band gap transition from the direct to indirect one 
occurs as the thickness for the 2D MoS$_2$ film increases from ML to bulk.~\cite{Chei2012}
Interestingly, when the thickness of 2D selenium increases from ML to TL, 
the location of the VBM moves from symmetry point $\Gamma$ to M along the $\Gamma$-M line. 
The CBM is located at a general $k$-point along the X-M symmetry line. Nevertheless, it gets closer to the M
point with the thickness [see Figs. 2(a), (b) and (c)]. Clearly, the band gap of 2D selenium 
is larger than that of the corresponding 2D tellurium (see Table III) due to the increased ionization energy. 
In general, the band gaps of 2D Se and Te materials are tunable by varying the number of layers.

It is well known that the band gap of a semiconductor is generally underestimated by the GGA calculations (see, e.g., Table III). Therefore,
to obtain more accurate band gaps, we also calculate the band structures by using the hybrid Heyd-Scuseria-Ernzerhof (HSE) functional \cite{heyd2003},
which is known to produce much improved band gaps for semiconductors. The theoretical band gaps 
from the HSE and GGA calculations together with the experimental
values are listed in Table III. The band structures from the HSE calculations with the spin-orbit coupling (SOC) 
included are displayed in Fig. S1. 
Indeed, Table III shows that the band gaps of bulk selenium and tellurium from the HSE-SOC calculations~\cite{Cheng2019} agree rather well 
with the corresponding experimental values.
Therefore, we use the band gaps from the HSE-SOC calculations and the scissors correction scheme~\cite{levine_prb_1991}
to evaluate linear and NLO properties for 2D Se and Te considered here.
We notice that the band gap from the HSE-SOC calculation is 1.128 (1.072) eV in BL (TL) tellurium, which agrees rather
well with that reported in Ref. \cite{qiao2018} (see Table III). The small difference stems from the fact 
that the band gaps from Ref. \cite{qiao2018} are
based on the optB88-vdW structure, while we use the structures from the SCAN structural optimizations. 
It is worthwhile to point out that the gap of BL Te from the HSE-SOC calculations is close to 1.3 eV, the ideal gap value 
for perfect solar energy absorption efficiency~\cite{Cai2017}.
Therefore, one could expect BL Te to have high solar energy absorption efficency,
thereby leading to high power conversion efficiency for photovoltaic devices.

We notice that the experimental band gaps for BL and TL tellurium are smaller than that from our HSE-SOC
calculations. This discrepancy is likely due to the effect of the graphene substrate because the environment
can significantly affect the electronic properties of 2D materials~\cite{Huser2013,Ugeda2014,Rasmussen2015}.
The graphene substrate can reduce the band gap of BL and TL tellurium mainly in two ways, namely, 
via (1) changing the structural parameters due to lattice mismatch and (2) offering extra dielectric screening. 
In particular, the in-plane lattice constants for atomically thin Te films on graphene substrate 
were reported~\cite{Huang2017} to be $a = 5.93$ \AA$ $ and $b = 4.42$ A, which are 2.3 \% and 3.3 \% 
larger than that of BL Te (see Table II), respectively. It was reported~\cite{Rasmussen2015} that
the band gap of MoS$_2$ ML would get reduced by as much as 0.46 eV upon a lattice expansion of 2 \%. 
Our GGA calculation for BL Te using the reported lattice constants~\cite{Huang2017} results in 
a band gap reduction of $\sim$0.1 eV. The substrates have also been found 
to introduce significant extra screening, thus reducing the band gap (see, e.g., Refs. ~\cite{Huser2013,Ugeda2014}). 
For example, putting a BN sheet on graphene would reduce the band gap of BN ML by 1.02 eV (14 \%)~\cite{Huser2013}. 
Nonetheless, further investigations into the reduction of the band gaps of 2D tellurium due to substrates 
are beyond the scope of this paper.

\section{Linear optical property}

\begin{figure}[htb]
\begin{center}
\includegraphics[width=8.0cm]{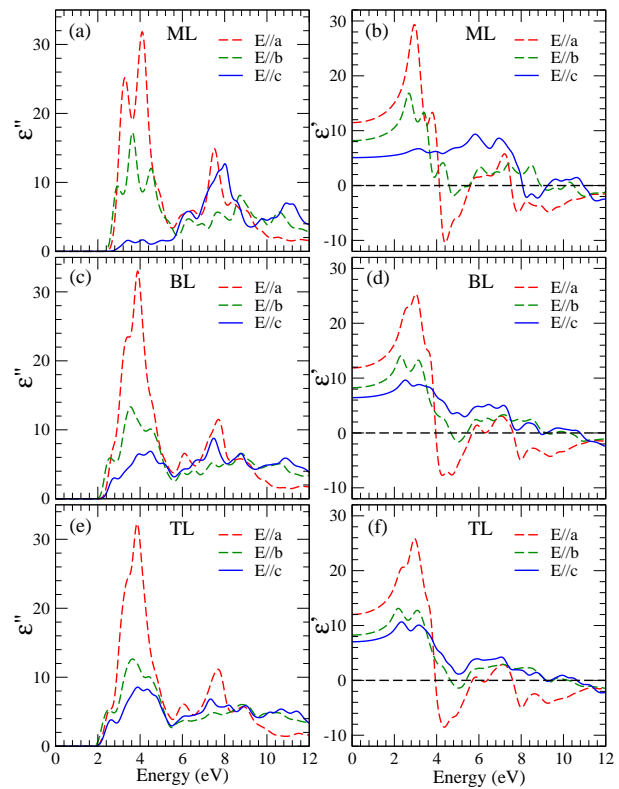}
\end{center}
\caption{The imaginary [$\varepsilon''(\omega)$]  and real part [$\varepsilon'(\omega)$] of the dielectric
function of (a, b) ML, (c, d) BL and (e, f) TL selenium
for both light polarization perpendicular (E$\parallel$a and E$\parallel$b) and parallel (E$\parallel$c) to the $c$ axis.
}
\end{figure}

In Fig. 3 and Fig. 4, the calculated real and imaginary parts
of the optical dielectric function $\varepsilon(\omega)$ of 2D selenium and tellurium are shown, respectively.
Bulk Se and Te have a trigonal lattice with the helical chains along the $c$-axis [see Figs. 1(a), 1(b) and 1(c)]
and thus exhibit a rather strong uniaxial optical anisotropy~\cite{Cheng2019} because of the weak inter-chain 
interaction. In contrast, 2D Se and Te form a 2D rectangular lattice with the helical chains lying along the $a$-axis (see Fig. 1),
and this may result in a strong in-plane optical anisotropy. Consequently, the real and imaginary parts of the dielectric
function of 2D Se and Te compose of three independent components, namely, light polarization parallel ($E \parallel c$)
and perpendicular ($E \parallel a$ and $E \parallel b$) to the $c$ axis. 

Figure 3 shows that overall, the dielectric functions of ML, BL and TL selenium look rather similar
due to the weak interlayer interaction. For example, there is a very prominant peak near 4.0 eV
in the $\varepsilon"(\omega)$ spectrum of $E \parallel a$ for all three multilayers. 
Also, in the low-energy region of about 2.5-5.5 eV, the optical anisotropy is strong in all three multilayers. 
In the high-energy region of about 5.5-10 eV, there is a prominent peak and two shoulders for $E \parallel a$ 
for all three structures. For $E \parallel b$ and $E \parallel c$, the amplitude of the peaks
 gradually decreases with the film thickness.  There are also some differences. 
For example, Fig. 3(a) shows that in ML Se, there is a rather sharp peak near 3.3 eV for
$E \parallel a$ and also near 8.0 eV for $E \parallel c$ which, however, appears as a weak shoulder and bulge, respectively, 
in the $\varepsilon"(\omega)$ spectra of BL Se [Fig. 3(c)] and TL Se [Fig. 3(e)]. 
This difference may be caused by a stronger quantum confinement in ML selenium. 
Furthermore, Fig. 3(a) indicates that optical absorption for $E \parallel c$ in 2.5-5.5 eV is very weak. 
Nevertheless, the absorption increases significantly with the film thickness [Fig. 3(c) and Fig. 3(e)]. 
Additionally, these peaks become oscillatory bulges in the structures, particularly in TL selenium. 
This could be explained by the fact that the interlayer interaction is weak but not negligible.
In particular, the real and imaginary parts of the optical dielectric function $\varepsilon(\omega)$ 
for BL and TL selenium are rather similar except the $E \parallel c$ spectrum.
The same phenomenon has been observed in Cr$_2$Ge$_2$Te$_6$ \cite{fang2018}. 
It is noted that the optical spectra for $E \parallel a$ (parallel to the chain direction) and for $E \parallel c$
(perpendicular to the layers) of TL selenium are nearly identical to the corresponding spectra of bulk selenium~\cite{Cheng2019}.
In other words, the optical properties of TL selenium are like that of quasi-1D systems.

\begin{figure}[htb]
\begin{center}
\includegraphics[width=8.0cm]{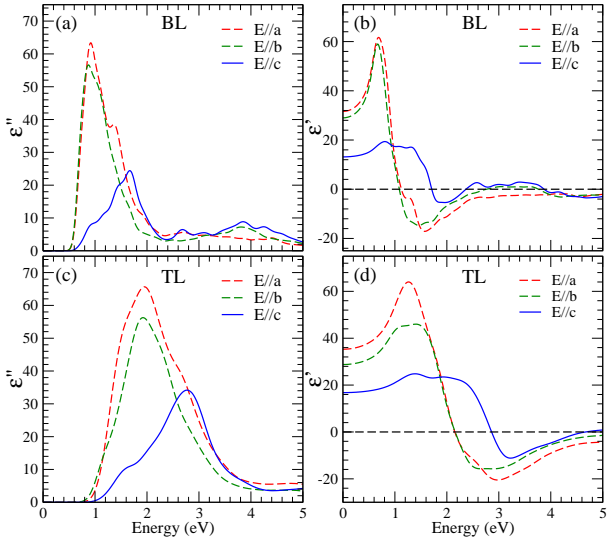}
\end{center}
\caption{The imaginary [$\varepsilon''(\omega)$] and real part [$\varepsilon'(\omega)$] of the dielectric
function of (a, b) BL and (c, d) TL tellurium
for both light polarization perpendicular (E$\parallel$a and E$\parallel$b) and parallel (E$\parallel$c) to the $c$ axis.
}
\end{figure}

As for 2D selenium, Fig. 4 indicates that overall, the dielectric functions 
of BL and TL tellurium are quite similar due to the weak interlayer interaction. 
In particular, there is a very prominant peak near 2.0 eV
in the $\varepsilon"(\omega)$ spectra of $E \parallel a$ and $E \parallel b$  for both multilayers.
Furthermore, the optical spectra of $E \parallel a$ and $E \parallel b$  for both multilayers 
are very similar, indicating almost no in-plane optical anisotropy.
This is in contrast to 2D selenium in which the peak height for $E \parallel b$ 
is much smaller than that of $E \parallel a$, indicating a strong in-plane optical anisotropy there. 
This could be explained by the fact that the relative in-plane inter-chain distance ($R/r$)
in 2D selenium is much larger than in 2D tellurium (see Table II).
On the other hand, Te 5$p$ orbitals are much larger than Se 4$p$ orbitals,
as indicated by the intrachain atomic distance ($r$) (see Table II).
Consequently, these result in a much stronger in-plane interchain bonding
in 2D tellurium and hence a much weaker in-plane optical anisotropy.
For $E \parallel c$, this peak shifts to $\sim$3.4 eV and $\sim$2.8  eV
for BL and TL tellurium [see Figs. 4(a) and 4(c)], respectively.
Also, in the low-energy region up to $\sim$3.0 eV, the optical spectra for in-plane and perpendicur 
polarizations differ significantly in both multilayers.     
Above 3.0 eV, the optical spectra for all three polarizations look rather similar.

\section{Second harmonic generation and linear electro-optic effect}

Bulk selenium and tellurium have five nonzero SHG susceptibility elements since their symmetry is either $D_3^4$ (right-handed screw) 
or $D_3^6$ (left-handed screw).~\cite{Cheng2019} However, 2D selenium and tellurium possess a lower symmetry 
and thus have more nonzero SHG susceptibility elements.
As mentioned above, their space group depends on the number of layers ($n$), namely,
$P2$ ($C_2^1$) for the odd $n$ and $P2_1$ ($C_2^2$) for the even $n$.
Nonetheless, the two space groups have the same nonzero elements. Therefore, all 2D selenium and tellurium materials
considered here have eight nonzero SHG elements as $\chi^{(2)}_{xxx}$,  $\chi^{(2)}_{yxy}$, 
$\chi^{(2)}_{xyy}$, $\chi^{(2)}_{zxy}$, $\chi^{(2)}_{xyz}$, $\chi^{(2)}_{yzx}$, $\chi^{(2)}_{zzx}$ and $\chi^{(2)}_{xzz}$.

\begin{table*}[htbp]
\begin{center}
\caption{
Selenium: Static dielectric constants ($\varepsilon_x, \varepsilon_y$, $\varepsilon_z$),
nonzero SHG susceptibility elements ($\chi^{(2)}_{xxx}$, $\chi^{(2)}_{xyy}$, $\chi^{(2)}_{xzz}$, 
$\chi^{(2)}_{xyz}$, $\chi^{(2)}_{yzx}$, $\chi^{(2)}_{yxy}$, $\chi^{(2)}_{zzx}$, $\chi^{(2)}_{zxy}$) (pm/V), 
and also nonzero LEO coefficients ($r_{xxx}$, $r_{xyy}$, $r_{xzz}$, $r_{xyz}$, $r_{yzx}$, $r_{yxy}$, $r_{zzx}$, $r_{zxy}$) (pm/V).
}
\begin{ruledtabular}
\begin{tabular}{p{0.5cm}p{0.5cm}p{0.4cm}p{0.6cm}p{0.5cm}p{0.5cm}p{0.5cm}p{0.5cm}p{0.5cm}p{0.5cm}p{0.5cm}p{0.5cm}p{0.9cm}p{0.7cm}p{0.7cm}p{0.9cm}p{0.7cm}p{0.7cm}p{0.7cm}p{0.7cm}}
       &$\varepsilon_x$ &$\varepsilon_y$ &$\varepsilon_z$ &$\chi^{(2)}_{xxx}$ &$\chi^{(2)}_{xyy}$ &$\chi^{(2)}_{xzz}$
                         &$\chi^{(2)}_{xyz}$ &$\chi^{(2)}_{yzx}$ &$\chi^{(2)}_{yxy}$ &$\chi^{(2)}_{zzx}$ &$\chi^{(2)}_{zxy}$
                         &$r_{xxx}$  &$r_{xyy}$ &$r_{xzz}$ &$r_{xyz}$ &$r_{yzx}$ &$r_{yxy}$ & $r_{zzx}$ &$r_{zxy}$ \\ \hline
 bulk  & 9.0\footnotemark[1] &   & 12.7\footnotemark[1] & 145\footnotemark[1] &  &  & 5\footnotemark[1]  &  &  & & & -3.58\footnotemark[1] &  &  & -0.12\footnotemark[1] &  &  &   &    \\
 ML    & 11.5 & 8.2 & 5.1 & 41 & 155 & 3 & 36 & 40 & 28 & 4 & 32 & -0.63 & -3.29 & -0.10 & -0.77 & -1.92 & -0.60 & -0.29 & -1.08  \\
 BL    & 11.9 & 8.2 & 6.4 & 17 & 92  & -6 & 27 & 27 & 12 & -11 & 25 & -0.24 & -1.87 & 0.15 & -0.54 & -1.01 & -0.24 & 0.52 & -0.65  \\
 TL    & 12.0 & 8.3 & 7.0 & 12 & 75 & -11 & 18 & 19 & 1 & -16 & 16 & -0.17 & -1.51 & 0.25 & -0.37 & -0.65 & -0.03 & 0.64 & -0.37     \\
\end{tabular}
\end{ruledtabular}
\footnotemark{Calculated value from reference~\onlinecite{Cheng2019}.}
\end{center}
\end{table*}

\begin{figure*}[htb]
\begin{center}
\includegraphics[width=16cm]{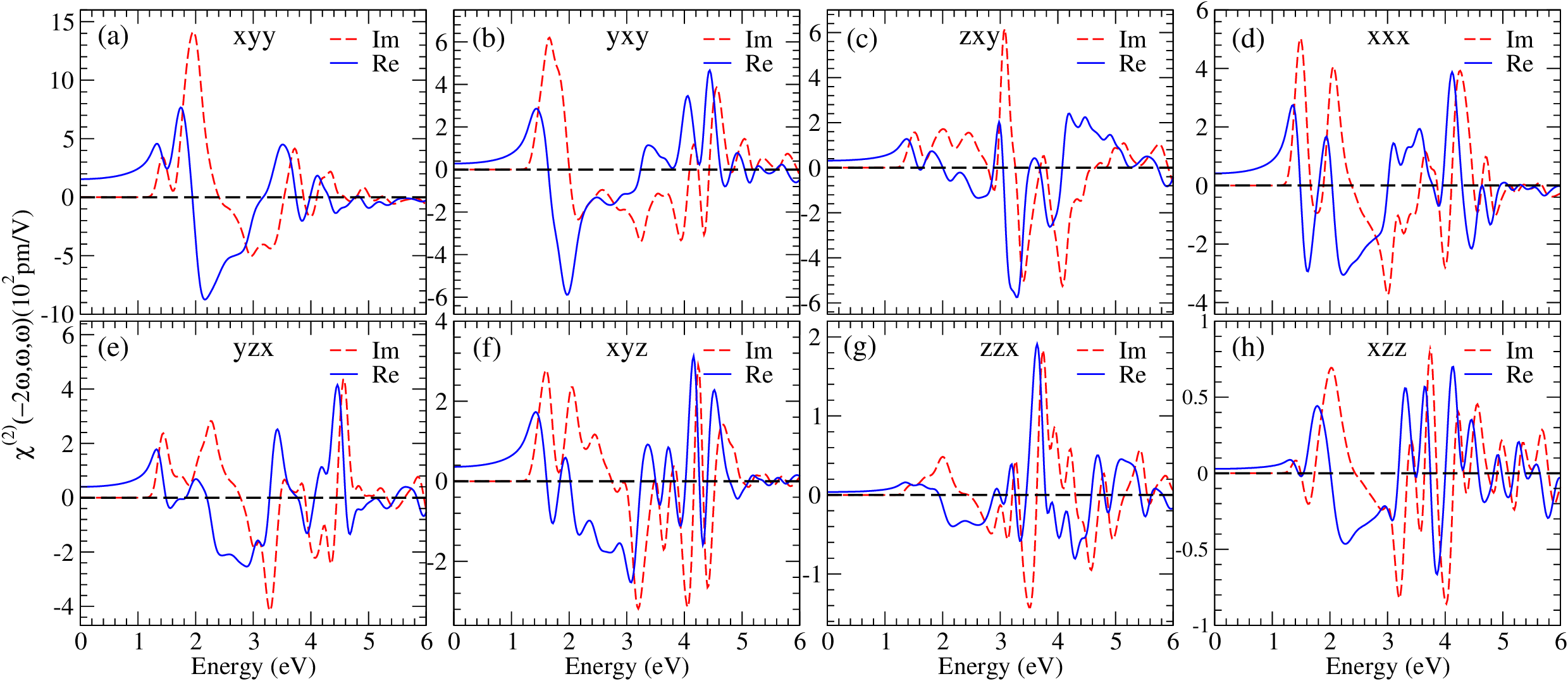}
\end{center}
\caption{Real and imaginary parts of the eight nonzero SHG susceptibility elements of ML selenium.}
\end{figure*}

\subsection{2D selenium}

\begin{figure*}[htb]
\begin{center}
\includegraphics[width=16cm]{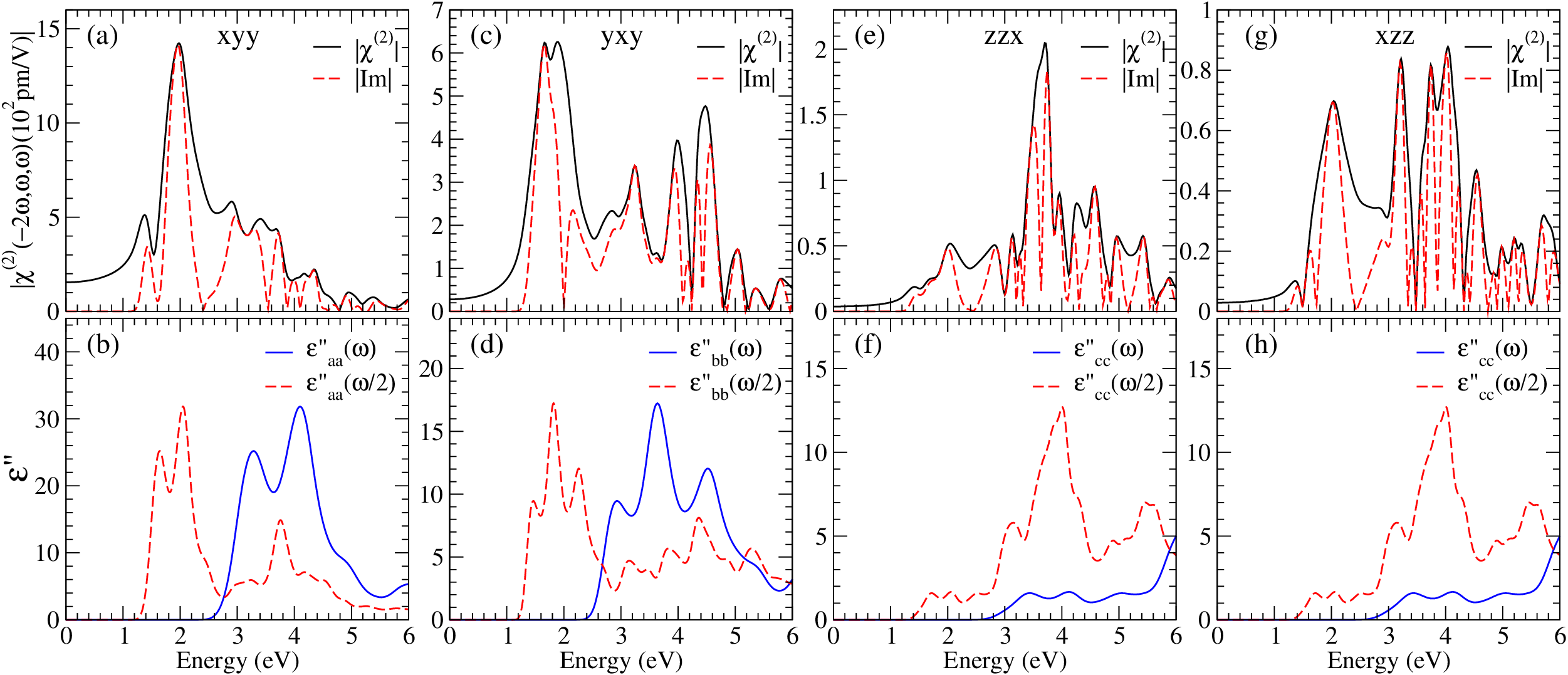}
\end{center}
\caption{Absolute value of nonzero SHG susceptibility elements (a) $\chi^{(2)}_{xyy}$, (c) $\chi^{(2)}_{yxy}$,
(e) $\chi^{(2)}_{zzx}$ and (g) $\chi^{(2)}_{xzz}$ of ML  selenium.
Imaginary part $\varepsilon''(\omega)$ of the dielectric function for light polarization
(b, d) perpendicular and (f, h) parallel to the $c$ axis.}
\end{figure*}

The real and imaginary parts of the SHG coefficients for ML selenium are displayed in Fig. 5,
while that of BL and TL selenium are shown, respectively, in Fig. S2 and Fig. S3 in the Supplementary Material (SM).~\cite{SM}
As for the linear optical property case, the calculated SHG coefficients for all the 2D Se structures are rather similar
(see Figs. 5, S2 and S3) due to the weak interlayer interaction.
For example, they all show that the $\chi^{(2)}_{xyy}$ spectra have the largest magnitude
among the nonzero SHG matrix elements in the entire optical frequency range, and their $\chi^{(2)}_{xyy}$ spectra
are similar to the $\chi^{(2)}_{xxx}$ spectra of bulk selenium~\cite{Cheng2019}.
They all have a prominent peak at 1.98 eV in the $\chi''^{(2)}_{xyy}$ spectrum. 
Nonetheless,  this peak height get reduced from ML Se ($\sim$1420 pm/V) to BL Se ($\sim$1020 pm/V) 
and to TL Se ($\sim$840 pm/V). The absolute value of SHG element $\chi^{(2)}_{xyy}$ of ML selenium at 1.98 eV
is 1423 pm/V, which is nearly 2 times larger than that of bulk selenium, and is also 6 times larger than
that of GaN \cite{Gavrilenko2000,Cai2009}, a widely used NLO semiconductor.
Figures 5, S2 and S3 indicate that all the nonzero SHG elements of all the 2D Se structures 
are rather large and purely dispersive for photon energy less than half of the band gap ($\frac{1}{2}E_g$).
For example, $\chi^{(2)}_{xyy}(0)$ of ML selenium (see Table IV) is 
nearly 20 times larger than $\chi^{(2)}_{zzz}(0)$ of GaN \cite{Gavrilenko2000,Cai2009}. 
This suggests that 2D selenium may have potential applications in low-loss NLO optical devices.

In what follows, we take ML selenium as an example to perform a detailed analysis.
In order to further analyze the prominent features in the calculated $\chi^{(2)} (\omega)$ spectra of ML selenium, 
we plot the absolute values of the imaginary part of $\chi^{(2)}$ and compare them 
with the absorptive part of the dielectric function $\varepsilon(\omega)$ in Fig. 6.
Because they have similar features, only four elements out of the eight nonzero SHG elements 
are presented in Fig. 6. The SHG involves not only single-photon $(\omega)$ resonance 
but also double-photon $(2\omega)$ resonance. Figure 6(a) shows
that the $\chi^{(2)}_{xyy}$ spectrum can be divided into two parts. The first part from 1.22 to 2.40 eV 
is primarily derived from double-photon resonances. On the other hand,
the second part (above 2.40 eV) originates predominantly from single-photon resonances 
with some contribution from double-photon resonances [see Figs. 6(a) and (b)].
In this energy region, the spectra oscillate rapidly and also decrease gradually with photon energy 
due to the mixing of one- and two-photon resonances. Notably,
for the $\vert\chi''^{(2)}_{yxy}\vert$ spectrum, the first prominent peak at $\sim$1.70 eV corresponds 
to the pronounced peak in the $\varepsilon_{bb}''(2\omega)$ [see Figs. 6(c) and 6(d)],
suggesting that it stems from two-photon resonances. It is clear from Fig. 6(d) that the peaks 
above the absorption edge of $\varepsilon_{bb}'' (\omega)$
are related to the $\varepsilon_{bb}'' (2\omega)$ and $\varepsilon_{bb}'' (\omega)$ spectra, 
indicating that they can be caused by both double- and single-photon resonances. 
Figures 6(e) and 6(g) show that the peaks in $\chi^{(2)}_{zzx}$ and $\chi^{(2)}_{xzz}$ from the absorption
edge of $\varepsilon_{cc}'' (2\omega)$ to the absorption edge of $\varepsilon_{cc}'' (\omega)$ may
be attributed to the two-photon resonances with E$\parallel$c [cf. $\varepsilon_{cc}'' (2\omega)$]. 
Above the absorption edge of $\varepsilon_{cc}'' (\omega)$, the absolute values of the imaginary part of 
$\chi^{(2)}(\omega)$ oscillate rapidly. This is consistent
with the fact that in this energy region $\chi^{(2)}(\omega)$ comes mainly from two-photon resonances 
with some contribution from one-photon resonances. Clearly, the 
modulus of the imaginary part of $\chi''^{(2)}_{xyy}$ is larger than that of $\chi''^{(2)}_{yxy}$,
and is also much larger than that of $\chi''^{(2)}_{zzx}$ and $\chi''^{(2)}_{xzz}$. 
Interestingly, this can be correlated with the magnitude relationship 
of $\varepsilon_{aa}'' (\omega)$, $\varepsilon_{bb}'' (\omega)$
and $\varepsilon_{cc}'' (\omega)$ [or $\varepsilon_{aa}'' (2\omega)$, $\varepsilon_{bb}'' (2\omega)$ 
and $\varepsilon_{cc}'' (2\omega)]$. 

The calculated static SHG susceptibility $\chi^{(2)}(0)$ and LEO coefficient $r(0)$ as well as 
dielectric constant $\varepsilon(0)$ of bulk and 2D selenium are presented in Table IV.
As for the optical frequency case discussed above, the static SHG susceptibility of 2D selenium
is also highly anisotropic. In particular, $\chi^{(2)}_{xyy}(0)$ of ML selenium is over 50 times larger than
$\chi^{(2)}_{xzz}(0)$ (see Table IV). Nevertheless, Table IV indicates that the anisotropy in the LEO coefficient $r(0)$
gets much reduced because of the strong anisotropy of the static $\varepsilon(0)$ [See Eq. (7)]. 
Remarkably, LEO coefficient $r_{xyy}$ ($\sim$ 3.29 pm/V) of ML Se 
is nearly 6 times larger than that of bulk GaN polytypes \cite{Gavrilenko2000,Cai2009},
suggesting that 2D selenium may become promising electric-optic materials for opto-electronic devices.

\begin{table*}[htbp]
\begin{center}
\caption{
Tellurium: Static dielectric constants ($\varepsilon_x, \varepsilon_y$, $\varepsilon_z$),
nonzero SHG susceptibility elements ($\chi^{(2)}_{xxx}$, $\chi^{(2)}_{xyy}$, $\chi^{(2)}_{xzz}$, 
$\chi^{(2)}_{xyz}$, $\chi^{(2)}_{yzx}$, $\chi^{(2)}_{yxy}$, $\chi^{(2)}_{zzx}$, $\chi^{(2)}_{zxy}$) (pm/V)
and also nonzero LEO coefficients ($r_{xxx}$, $r_{xyy}$, $r_{xzz}$, $r_{xyz}$, $r_{yzx}$, $r_{yxy}$, $r_{zzx}$, $r_{zxy}$) (pm/V).
}
\begin{ruledtabular}
\begin{tabular}{p{0.5cm}p{0.5cm}p{0.4cm}p{0.6cm}p{0.5cm}p{0.5cm}p{0.5cm}p{0.5cm}p{0.5cm}p{0.5cm}p{0.5cm}p{0.5cm}p{0.9cm}p{0.7cm}p{0.7cm}p{0.9cm}p{0.7cm}p{0.7cm}p{0.7cm}p{0.7cm}}
         &$\varepsilon_x$ &$\varepsilon_y$ &$\varepsilon_z$ &$\chi^{(2)}_{xxx}$ &$\chi^{(2)}_{xyy}$ &$\chi^{(2)}_{xzz}$
                         &$\chi^{(2)}_{xyz}$ &$\chi^{(2)}_{yzx}$ &$\chi^{(2)}_{yxy}$ &$\chi^{(2)}_{zzx}$ &$\chi^{(2)}_{zxy}$
                         &$r_{xxx}$  &$r_{xyy}$ &$r_{xzz}$ &$r_{xyz}$ &$r_{yzx}$ &$r_{yxy}$ & $r_{zzx}$ &$r_{zxy}$ \\ \hline
 bulk  & 33.2\footnotemark[1]  &   & 49.0\footnotemark[1] & 169\footnotemark[1] &  &   & 1009\footnotemark[1] &  &  &  &  & -0.30\footnotemark[1] &  &  & -1.82\footnotemark[1] &   &   &  &  \\
 BL    & 31.6 & 29.0 & 13.2 & 557 & 1077  & 31 & 592 & 585 & 897 & 0 & 591 & -1.11 & -2.34 & -0.15 & -1.29 & -3.06 & -1.95 & 0 & -2.84  \\
 TL    & 35.3 & 28.8 & 16.8 & 823 & 872 & 18 & 656 & 644 & 683 & -11 & 626 & -1.32 & -1.72 & -0.06 & -1.29 & -2.67 & -1.34 & 0.08 & -2.11      \\
\end{tabular}
\end{ruledtabular}
\footnotemark[1]{Calculated value from reference~\onlinecite{Cheng2019}.}
\end{center}
\end{table*}

\subsection{2D tellurium}
The calculated static dielectric constant $\varepsilon(0)$, SHG susceptibility $\chi^{(2)}(0)$ 
and LEO coefficient $r(0)$ of 2D and bulk tellurium are listed in Table V. 
First of all, we notice that 2D Te exhibit much stronger static SHG responses than 2D Se.
In particular, $\chi^{(2)}_{xyy}(0)$ for BL Te (Table V) is more than 100 times larger 
than that of GaN in both zinc-blende and wurtzite structures~\cite{Gavrilenko2000,Cai2009}. 
Thus, our calculations indicate that 2D Te may have valuable NLO applications such as frequency doublers. 
Nevertheless, as for 2D selenium, LEO coefficient ($r_{yzx}$) of BL Te is only 5 times greater than 
that of GaN \cite{Gavrilenko2000,Cai2009},
because static dielectric constant $\varepsilon(0)$ of BL Te also becomes larger than BL Se
[see Tables IV and V as well as Eq. (7)]. 

The calculated real and imaginary parts of the SHG susceptibility of TL and BL tellurium are displayed, respectively,
in Fig. 7 and Fig. S4 in the SM~\cite{SM}. Like 2D selenium, the spectral shape of the real and imaginary parts 
of $\chi^{(2)}(\omega)$ for BL and TL tellurium are rather similar, indicating a weak interlayer interaction. 
Nevertheless, unlike 2D selenium, the $\chi^{(2)}_{xxx}(\omega)$ spectrum has the largest magnitude in 2D tellurium (see Fig. 7).
Furthermore, when one moves from BL tellurium (Fig. S4) to TL tellurium (Fig. 7),
the magnitude of the $\chi^{(2)}_{xxx}(\omega)$ spectrum increases significantly rather than decreases.
Another striking difference is that the $\chi^{(2)}(\omega)$ spectra of 2D tellurium in general
are more than 10 times larger than the corresponding spectra in 2D selenium (see, e.g., Figs. 7 and S3). 
In particular, the peak height near 1.2 eV in the $\chi''^{(2)}_{xxx}(\omega)$ spectrum of TL selenium
[Fig. S3(g)] more than 40 times smaller than that of TL tellurium [Fig. 7(a)]. 
For a specific semiconductor, the smaller the band gap is, the larger the magnitude of the second-order NLO susceptibility, 
as can be seen from Eqs. (4) and (5).
This is the main reason why 2D tellurium generally have much larger SHG coefficients than 2D selenium
(see also Tables IV and  V) because of the much smaller band gaps of 2D tellurium.

\begin{figure*}[htb]
\begin{center}
\includegraphics[width=16cm]{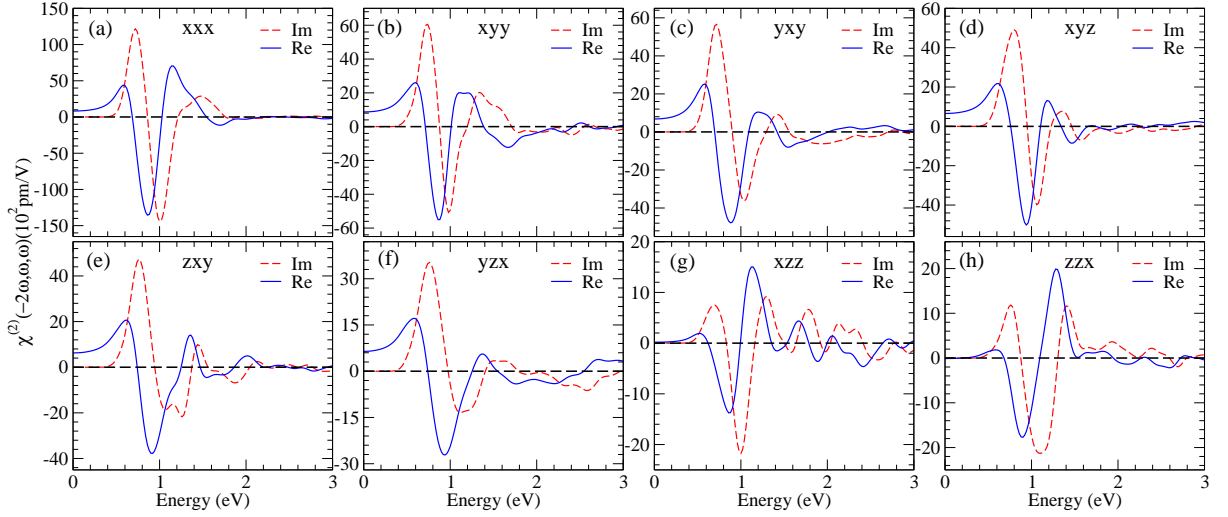}
\end{center}
\caption{Real and imaginary parts of the eight nonzero SHG susceptibility elements of TL tellurium.}
\end{figure*}

Let us now focus on the $\chi^{(2)}_{xxx}$ spectrum since it is the largest nonzero element 
among all the nonvanishing SHG coefficients of 2D tellurium. Figure 7(a) shows that 
$\chi^{(2)}_{xxx}$ is large in the photon energy range from 0.0 eV to $\sim$1.86 eV. 
In the energy range from 0.0 to 0.22 eV, the real part of the
$\chi^{(2)}_{xxx}$ remains almost constant. As photon energy further increases, 
it increases steadily, and produces a small peak at $\sim$0.58 eV [see Fig. 7(a)].
In the energy region of 0.70-1.02 eV, the spectrum of $\chi'^{(2)}_{xxx}$  
becomes negative and forms a huge negative peak at $\sim$0.86 eV. Beyond 1.02 eV,
it becomes positive again and decreases gradually in the higher energy region. 
The imaginary part of $\chi^{(2)}_{xxx}$ forms a sine-function shape in the photon energy
range from $\sim$0.5 to 1.7 eV with a large maximum of 12000 pm/V at $\sim$0.7 eV
and a large minimum of $-14000$ pm/V at $\sim$1.0 eV [see Fig. 7(a)]. 
It is negligible outside this photon energy range.
Interestingly, all the nonzero SHG coefficients of TL tellurium have nearly identical shapes (see Fig. 7).
Finally, we note that the magnitude of $\vert$$\chi^{(2)}_{xxx} (-2\omega,\omega,\omega)$$\vert$
at 0.98 eV is as high as 15000 pm/V, which is 4 times larger than that of bulk Te \cite{Cheng2019} 
and 65 times larger than that of GaN \cite{Gavrilenko2000,Cai2009}. 
All these suggest that few-layer tellurium would find promising applications in, e.g., 
optical switching, ultrathin second-harmonic and sum frequency generation devices, 
optical modulation, and light signal modulators.

\begin{figure*}[htb]
\begin{center}
\includegraphics[width=16cm]{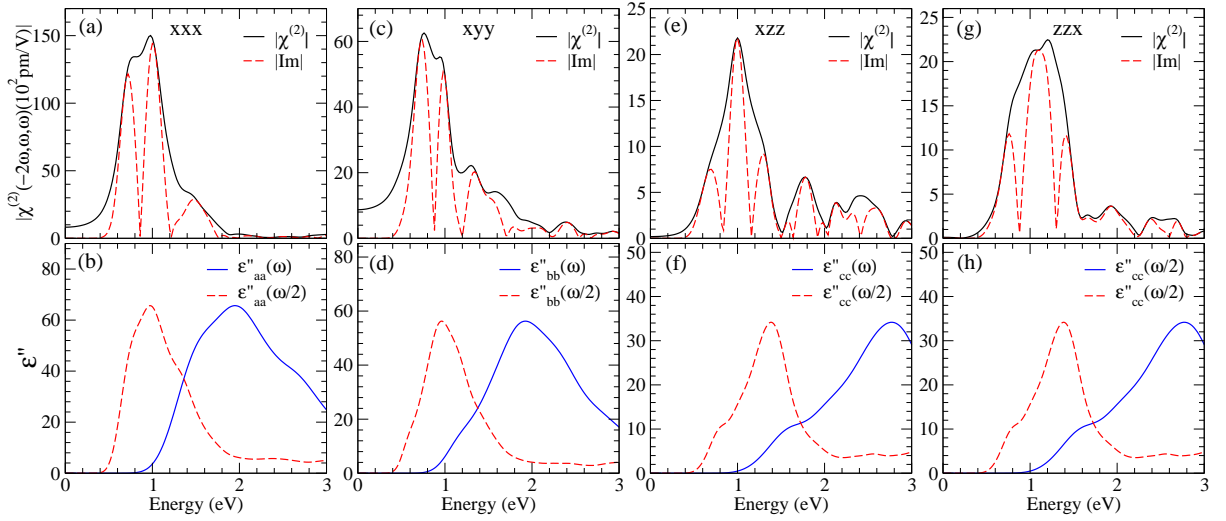}
\end{center}
\caption{Absolute values of nonzero SHG susceptibility elements (a) $\chi^{(2)}_{xxx}$, (c) $\chi^{(2)}_{xyy}$,
(e) $\chi^{(2)}_{xzz}$ and (g) $\chi^{(2)}_{zzx}$ of TL tellurium.
Imaginary part $\varepsilon''(\omega)$ of the dielectric function for light polarization
(b, d) perpendicular and (f,h) parallel to the $c$ axis.}
\end{figure*}

In order to further understand the features in the calculated SHG spectra, the absolute values of the imaginary parts 
of four prominent elements $\chi^{(2)}_{xxx}$, $\chi^{(2)}_{xyy}$,  $\chi^{(2)}_{xzz}$ and  $\chi^{(2)}_{zzx}$ of TL tellurium 
are plotted in Fig. 8 as examples, together with the imaginary part of the dielectric function. 
Figure 8 suggests that the spectral feature of both $\chi^{(2)}_{xzz}$ and $\chi^{(2)}_{zzx}$
between $\sim$0.44 and $\sim$1.72 eV could be attributed to the two-photon resonances shown in $\varepsilon_{cc}'' (2\omega)$. 
In contrast, the structure above $\sim$1.72 eV stems from both single-photon [see $\varepsilon_{cc}'' (\omega)$] 
and two-photon resonances [see $\varepsilon_{cc}'' (2\omega)$].
As a result, the SHG spectra, especially $\chi^{(2)}_{xzz}$, oscillate rapidly and diminish gradually 
in the higher energy range [see Figs. 8(e) and 8(g)].
Similarly, for $\chi^{(2)}_{xxx}$ and $\chi^{(2)}_{xyy}$, the spectral structures between $\sim$0.34 eV and $\sim$1.36 eV are mainly
due to double-photon ($2\omega$) resonances for $E\parallel a$ and $E\parallel b$ 
[see $\varepsilon_{aa}'' (2\omega)$ and $\varepsilon_{bb}'' (2\omega)$], respectively.
The features in $\chi^{(2)}_{xxx}$ and $\chi^{(2)}_{xyy}$ for photon energies above 1.36 eV could be attributed 
to the single-photon resonances as can be seen from $\varepsilon_{aa}'' (\omega)$ and
$\varepsilon_{bb}'' (\omega)$, respectively. Clearly, the magnitude of $\chi^{(2)}_{xxx}$ is largest, and also is 
about several times larger than the other nonzero elements (see Fig. 7). Interestingly,
this seems to be correlated with the fact that the magnitude of $\varepsilon_{aa}'' (\omega)$
[$\varepsilon_{aa}'' (2\omega)$] is larger than that of $\varepsilon_{bb}'' (\omega)$ [$\varepsilon_{bb}'' (2\omega)$], 
and much larger than $\varepsilon_{cc}'' (\omega)$ [$\varepsilon_{cc}'' (2\omega)$ ]. 

\section{Bulk Photovoltaic Effect}

\begin{figure}[htb]
\begin{center}
\includegraphics[width=7cm]{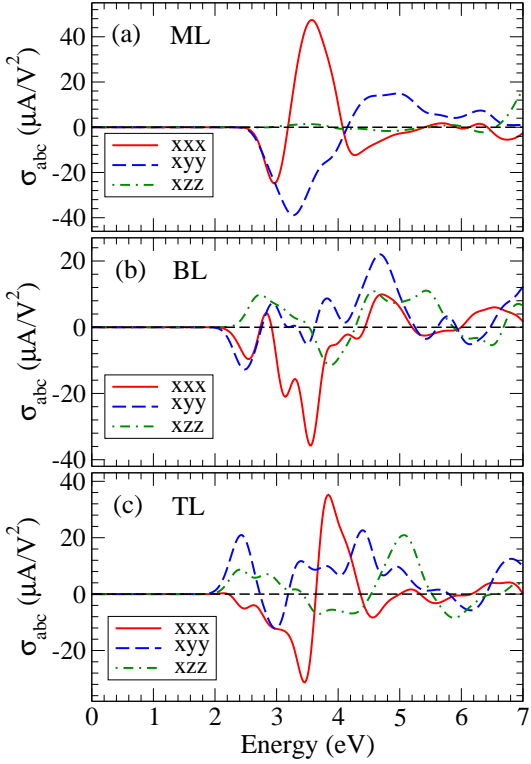}
\end{center}
\caption{Shift current conductivity ($\sigma$) versus photon energy for (a) bulk selenium and (b) bulk tellurium.}
\end{figure}

The generation of photocurrents is a crucial component for solar energy harvesting.
A promising mechanism for the photocurrent generation is the BPVE in which 
DC currents are generated in a noncentrosymmetric material under light irradiation.
Young {\it et al.} recently demonstrated that the shift current dominates the BPVE in BaTiO$_3$~\cite{Young2012}.
In this section, we present the calculated nonzero elements of the shift current conductivity tensor 
for all the 2D selenium and tellurium considered here.

We notice that nonzero elements of the shift current conductivity tensor are 
the same as that of the SHG susceptibility tensor.
Therefore, for all the 2D selenium and tellurium considered here,
the nonzero elements are 
$\sigma_{xxx}$, $\sigma_{yxy}$, $\sigma_{xyy}$, $\sigma_{zxy}$, $\sigma_{xyz}$, $\sigma_{yzx}$, $\sigma_{zzx}$ and $\sigma_{xzz}$. 
However, since we consider only the shift current generation due to the linearly polarized light here,
we will present only the calculated shift current elements of $\sigma_{xxx}$, $\sigma_{xyy}$  and $\sigma_{xzz}$ in this paper. 
Furthermore, here we have also calculated the shift current conductivity tensors 
for bulk selenium and bulk tellurium for comparison. The computational details were
already reported in Ref. ~\cite{Cheng2019}. Their space group is $P3_121$ and there are
two independent shift current elements among the five nonzero elements, i.e.,
$\sigma_{xxx} = -\sigma_{xyy}=-\sigma_{yxy}$ and $\sigma_{xyz} = -\sigma_{yzx}$.
The calculated two independent shift current elements of bulk selenium and bulk tellurium 
are displayed in Fig. 9.

\begin{figure}[htb]
\begin{center}
\includegraphics[width=7cm]{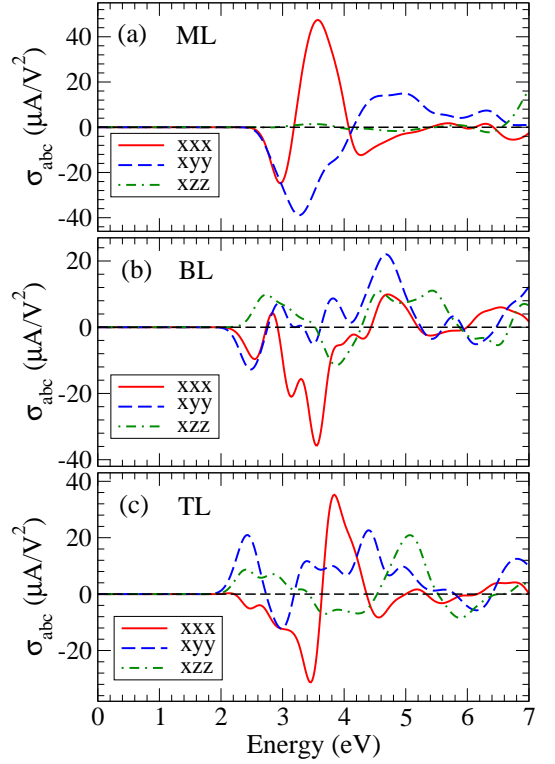}
\end{center}
\caption{Shift current conductivity ($\sigma$) versus photon energy for (a) ML, (b) BL and (c) TL selenium.}
\end{figure}

\subsection{2D selenium}
The calculated shift current $\sigma_{abc}$ spectra of 2D selenium are plotted in Fig. 10. 
It is clear that unlike the dielectric function and SHG spectra, the shift current spectra exhibit 
a strong layer-thickness dependence. Furthermore, the $\sigma_{xxx}$ spectrum in general has the largest
magnitude and the $\sigma_{xzz}$ has the smallest magnetiude. In particular, the $\sigma_{xzz}$ spectrum
of ML selenium is negligibly small compared with the $\sigma_{xxx}$ and $\sigma_{xyy}$ spectra [see Fig. 10(a)].
This is clearly due to the very small absorption of light polarized along the layer normal direction
in this photon energy range [see Fig. 3(a)]. As the layer thickness increases when one moves
from ML selenium to BL and TL selenium, the optical absorption of the out-of-plane polarized light 
increases (see Fig. 3) and hence the magnitude of the $\sigma_{xzz}$ spectrum also increases (see Fig. 10). 
Interestingly, the height (47 $\mu$A/V$^2$) of the maximal peak of the $\sigma_{xxx}$ spectrum 
at 3.58 eV of ML selenium is larger than that of the $\sigma_{xxx}$ spectrum at 3.56 eV (-36 $\mu$A/V$^2$) 
for BL selenium and also at 3.84 eV (35 $\mu$A/V$^2$) for TL selenium.
We notice that the magnitudes of the $\sigma_{xxx}$ spectra of 2D selenium are at least
comparable to that of the $\sigma_{zxx}$ and $\sigma_{zzz}$ spectra of BaTiO$_3$ \cite{Young2012}, 
an archetypical ferroelectric BPVE material. 
These magnitudes are also comparable to that of ML group-IV monochalcogenides (e.g., GeS)~\cite{Rangel2017}. 
Therefore, 
2D selenium may become excellent materials for photovoltaic solar cells. 
Figures 9(a) and 10 indicate that the shift current conductivities  of bulk and 2D selenium
are in the same order of magnitude, although their spectral shapes are quite different. 

\begin{figure}[htb]
\begin{center}
\includegraphics[width=7cm]{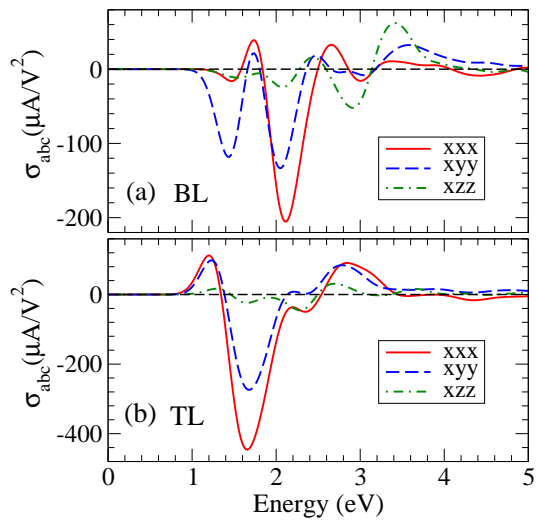}
\end{center}
\caption{Shift current conductivity ($\sigma$) versus photon energy for (a) BL and (b) TL tellurium.
}
\end{figure}

\subsection{2D tellurium}
We display the calculated shift current conductivity ($\sigma_{abc}$) spectra of BL and TL tellurium in Fig. 11.          
First of all, we notice that the $\sigma_{abc}$ spectra of 2D tellurium are large in the optical frequency range,
being nearly one-order of magnitude larger than the corresponding spectra of 2D selenium (see Fig. 10).
This notable difference between 2D tellurium and 2D selenium may be attributed to the smaller band gaps
and highly covalent bonding in 2D tellurium.
Furthermore, the magnitude of the $\sigma_{xxx}$ at 1.66 eV of TL tellurium is nearly 4 times larger than
the maximum of the shift current conductivities of ML GeS~\cite{Rangel2017}.
In fact, the shift current conductivity of ML Ge and Sn monochalcogenides apparently is larger than that of 
other noncentrosymmetric semiconductors reported so far~\cite{Rangel2017}.
This indicates that 2D tellurium are very promising materials for photovoltaic devices.
The $\sigma_{xxx}$ spectrum in general has the largest
magnitude while the $\sigma_{xzz}$ has the smallest magnetiude.
Indeed, the $\sigma_{xzz}$ spectrum of TL tellurium is negligibly small 
compared with the $\sigma_{xxx}$ and $\sigma_{xyy}$ spectra [see Fig. 11(b)].
Second, unlike the dielectric function and SHG spectra, the shift current spectra exhibit
a strong layer-thickness dependence. 
In particular, the height of the prominent negative peak at $\sim$1.7 eV in the $\sigma_{xzz}$ spectrum
of TL tellurium is nearly twice as large as that of the negative peak at $\sim$2.1 eV in BL tellurium (see Fig. 11). 
In contrast to 2D selenium, 2D tellurium have the shift current conductivities that 
are much larger than bulk tellurium (see Figs. 9(b) and 11). In particular, the shift current conductivities 
of TL Te [Fig. 11(b)] are as much as five times larger than that of bulk tellurium [Fig. 9(b)].
Our work thus demonstrates that few-layer tellurium with narrow band gaps
are promising materials for high efficient photovoltaic solar cells.

\begin{figure*}[htb]
\begin{center}
\includegraphics[width=14cm]{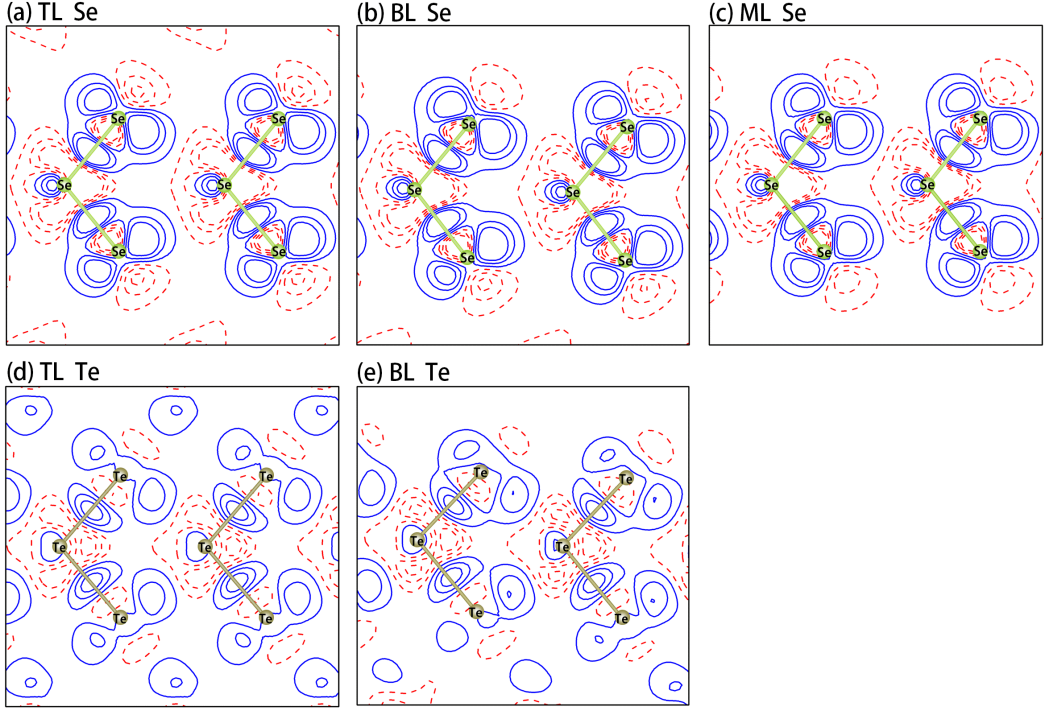}
\end{center}
\caption{Contour plots of the deformation charge density in (a) TL, (b) BL and (c) ML selenium
as well as in (d) TL and (e) BL tellurium. The contour interval is 0.02 e/\AA$^3$.
Positive contours (blue solid lines) indicate electron accumulation and negative contours (red dashed lines)
depict electron depletion. }
\end{figure*}

\section{Discussion}
For a specific semiconductor, low frequency NLO properties depend significantly on its band gap. 
This could be seen from the fact that the energy differences between the initial and final states of the optical transitions 
are in the denominators of Eqs. (4), (5) and (9). This indicates that the smaller the band gap is, 
the larger the magnitude of the NLO responses would be. For example, for the imaginary part of the SHG coefficients,
the magnitude in the low frequency region would be roughly proportional to the inverse of the fourth power of the band gap
[see Eqs. (4) and (5)]. 
Equation (9) also indicates that the magnitude of the shift current conductivity is roughly
proportional to the inverse of the third power of the band gap.
As a result, although 2D Se and Te have identical crystalline structures and similar band structures,
the calculated imaginary part of the dielectric function, SHG coefficients and shift photoconductivities for
2D Te are much larger than that of 2D Se simply because 2D Te has a much smaller band gap. 
Interestingly, unlike bulk Te, few-layer Te have a band gap being larger than the optical frequencies of interest, 
showing that the materials would be useful for the NLO applications. 

As mentioned in the preceding sections, compared to other semiconductors with similar band gaps, 
2D Se and Te generally exhibit much larger NLO effects. 
To further understand the origins of the enhanced NLO responses of 2D selenium and tellurium, 
we calculate the difference charge density which is defined as the difference
between the total charge density and the superposition of the free atomic charge densities, as displayed in Fig. 12. 
Figures 12(a), 12(b) and 12(c) [Figs. 12(d) and 12(e)] show that a considerable electron charge buildup occurs 
in the vicinity of the Se-Se (Te-Te) bond center by depleting the charge around the atoms along the bond directions in 2D selenium (2D tellurium). 
This indicates that there is strong directional covalent bonding in these 2D materials. 
It is known that strong covalent bonding would generate large spatial overlap between the wave functions
of initial and final states and thus leads to large optical matrix elements, thereby resulting in enhanced NLO responses. 
Furthermore, Fig. 12 also shows a charge buildup around each atom in the direction
perpendicular to the chain, indicating the existence of lone-pair electrons. 
The presence of lone-pair electrons is beneficial to the generation of induced dipole oscillations by the optical electric fields, 
thus leading to larger $\chi^{(2)}$ values \cite{Jiang2014,Cammarata2014} and other NLO effects. 
We notice that these features in the difference charge density distributions in 2D Se and Te
are quite similar to that found in bulk trigonal Se and Te~\cite{Cheng2019}. 
Bulk Se and Te crystalize in helical chainlike structures and also exhibit large SHG and LEO effect~\cite{Cheng2019}.
Therefore, in general, quasi one-dimensional crystals with strong directional covalency and lone-pair
electrons would possess large NLO values. Moreover, high anisotropy would result in large joint DOS, 
and hence give rise to larger BPV and SHG effect \cite{Ingers1988,Song2009}.

Therefore, in general, to search for the materials with large NLO responses, one first should focus on those semiconductors with smallest
possible band gaps which are larger than the optical frequencies required by specific NLO applications.
Second, one should also pay attention to low dimensional systems with 
high anisotropy, strong covalency and/or lone-pair electrons.

\section{CONCLUSION}
We have carried out a systematic first-principles investigation on the nonlinear optical properties of 2D selenium and tellurium within
the generalized gradient approximation plus scissors correction.
We find that all the 2D materials possess large NLO responses. In particular, TL Te exhibits a SHG coefficient $\chi^{(2)}_{xxx}$
of up to 15000 pm/V, which is 65 times larger than that of bulk GaN. 
BL Te exhibits gigantic static SHG coefficient $\chi^{(2)}_{xyy}$,
which is more than 100 times larger than that of bulk GaN. 
ML Se also displays prominent NLO effects with maximal nonvanishing SHG coefficient $\chi^{(2)}_{xyy}$ being as large as $\sim$1400 pm/V,
which exceeds 6 times larger than that of GaN. 
Furthermore, ML Se and BL Te have large LEO coefficients $r_{xyy}$ and $r_{yzx}$,
being up to 6 times and 5 times larger than that of bulk GaN polytypes, respectively. 
We also find that the maximum of the shift current in the visible frequency range 
for TL Te is two times larger than that of ML GeS.
All these findings suggest that 2D selenium and tellurium are promising NLO materials for applications 
in, e.g., photovoltaic devices, second-harmonic generation, sum frequency generation, frequency conversion,
phase matching, electro-optical switches and light signal modulators. 
The features in the SHG spectra of the 2D Se and Te
are correlated with the peaks in the imaginary part of the corresponding optical dielectric function 
in terms of single-photon and double-photon resonances.
We believe that this work will stimulate further experiments on the second-order nonlinear optical responses of these fascinating elemental 2D materials.

\section*{ACKNOWLEDGEMENTS}

M. C. thanks Department of Physics and Center for Theoretical Physics, National Taiwan University
for its hospitality during her three months visit.
Work at Xiamen University is supported by the National Key R$\&$D Program of China (Grant No. 2016YFA0202601),
and the National Natural Science Foundation of China (No. 11574257).
G. Y. G. acknowledges support from the Ministry of Science and Technology, the Academia Sinica,
the National Center for Theoretical Sciences in Taiwan.

\end{document}